\documentclass[sigconf]{acmart}
\AtBeginDocument{%
  }

\renewcommand\footnotetextcopyrightpermission[1]{}

\usepackage{alltt}
\usepackage{microtype}
\usepackage{graphicx}
\usepackage{subcaption}
\usepackage{booktabs} % for professional tables
\usepackage{multirow}
\usepackage{soul}
\include{todonotes}
\usepackage{hyperref}
\usepackage{tikz}
\usepackage{enumitem}
\usepackage{balance}

\usepackage{xcolor}
\usepackage{natbib}
\usepackage{graphicx}
\usepackage{listings}
\usepackage{subcaption}

\definecolor{background}{HTML}{F7F7F7}
\definecolor{keyword}{HTML}{37AC4A} 
\definecolor{operator}{HTML}{A51DFF}
\definecolor{string}{HTML}{C03333}

\lstdefinelanguage{python}{
  stringstyle=\textcolor{string},
  showstringspaces=false,
  morestring=[b]",
  morestring=[b]',
  morestring=[b]""",
  morecomment=[l]\#,
  morekeywords={and,as,assert,break,class,continue,def,del,elif,else,except,False,finally,for,from,global,if,import,in,is,lambda,None,nonlocal,not,or,pass,raise,return,True,try,while,with,yield},
  keywordstyle=\color{keyword}\bf\sffamily,
  commentstyle=\color{gray}\bf\sffamily,
  literate=
    *{>>}{{\bf\texttt{\color{operator}{>{}>}}}}{1}
    {\&}{{\bf\texttt{\color{operator}{\&}}}}{1}
    {|}{{\bf\texttt{\color{operator}{|}}}}{1}
    {=}{{\bf\texttt{\color{operator}{=}}}}{1}
    {(}{{\bf\texttt{\color{keyword}{(}}}}{1}
    {)}{{\bf\texttt{\color{keyword}{)}}}}{1}
    {[}{{\bf\texttt{\color{keyword}{[}}}}{1}
    {]}{{\bf\texttt{\color{keyword}{]}}}}{1}
    {\{}{{\bf\texttt{\color{keyword}{\char '173}}}}{1}
    {\}}{{\bf\texttt{\color{keyword}{\char '175}}}}{1},
}

\usepackage{pifont}

\usepackage{cuted}
\definecolor{Gray}{gray}{0.3}
\tikzstyle{mybox} = [draw=black, very thick, rectangle, rounded corners, inner ysep=5pt, inner xsep=5pt, fill=gray!20]

\newcommand*{\failtopass}[0]{\mbox{$F\!\!\to\!\!P$}\xspace}

\newcommand{\findings}[2]{
    \smallskip
    \noindent
    \begin{tikzpicture}
        \node [mybox] (box){%
        \centering
        \begin{minipage}{.95\columnwidth}
        \fontsize{8.8}{10}\selectfont
        \textbf{Finding #1}. #2
        \end{minipage}
        };
    \end{tikzpicture}%
}

\usepackage{xspace}

\usepackage[capitalize,noabbrev]{cleveref}

\begin{document}

\title{Reproduction Test Generation for Java SWE Issues}


\begin{comment}
\author{Toufique Ahmed, Jatin Ganhotra, Avraham Shinnar, and Martin Hirzel}
\affiliation{%
  \institution{IBM}
  \city{Yorktown Heights}
  \state{New York}
  \country{USA}
  \postcode{10603}}
\email{tfahmed@ibm.com, {jatinganhotra, shinnar, hirzel}@us.ibm.com}

\end{comment}

\author{Toufique Ahmed}
\affiliation{%
  \institution{IBM}
  \city{Yorktown Heights}
  \state{New York}
  \country{USA}
  \postcode{10603}}
\email{tfahmed@ibm.com}

\author{Jatin Ganhotra}
\affiliation{%
  \institution{IBM}
  \city{Yorktown Heights}
  \state{New York}
  \country{USA}
  \postcode{10603}}
\email{jatinganhotra@us.ibm.com}

\author{Avraham Shinnar}
\affiliation{%
  \institution{IBM}
  \city{Yorktown Heights}
  \state{New York}
  \country{USA}
  \postcode{10603}}
\email{shinnar@us.ibm.com}

\author{Martin Hirzel}
\affiliation{%
  \institution{IBM}
  \city{Yorktown Heights}
  \state{New York}
  \country{USA}
  \postcode{10603}}
\email{hirzel@us.ibm.com}

\begin{abstract}
Given an issue on a software repository, a reproduction test confirms
its presence in the code before it gets fixed and its absence after.
Reproduction tests provide crucial execution-based feedback for
diagnosis and validation during software development.
Unfortunately, they are usually missing.
Therefore, recent work has introduced both benchmarks and a thriving
literature on solutions for reproduction test generation from issues.
However, that work has focused on Python and neglected other
languages such as Java, which is important for enterprise software.
This paper introduces both a benchmark and a solution for Java
repository-level reproduction test generation.
The benchmark, TDD-Bench-Java, is the first to model this problem and
comprises 250 instances sourced from popular open-source repositories.
The solution, e-Otter++ for Java, adapts a state-of-the-art reproduction
test generator for Python to yield high performance on Java.
To evaluate in an industry setting, besides empirical results with
TDD-Bench-Java, this paper also presents results with a
contamination-free proprietary dataset.
Overall, we hope that this paper contributes to bringing better
diagnosis and validation to Java software development.

\end{abstract}


\begin{comment}
\begin{CCSXML}
<ccs2012>
 <concept>
  <concept_id>00000000.0000000.0000000</concept_id>
  <concept_desc>Do Not Use This Code, Generate the Correct Terms for Your Paper</concept_desc>
  <concept_significance>500</concept_significance>
 </concept>
 <concept>
  <concept_id>00000000.00000000.00000000</concept_id>
  <concept_desc>Do Not Use This Code, Generate the Correct Terms for Your Paper</concept_desc>
  <concept_significance>300</concept_significance>
 </concept>
 <concept>
  <concept_id>00000000.00000000.00000000</concept_id>
  <concept_desc>Do Not Use This Code, Generate the Correct Terms for Your Paper</concept_desc>
  <concept_significance>100</concept_significance>
 </concept>
 <concept>
  <concept_id>00000000.00000000.00000000</concept_id>
  <concept_desc>Do Not Use This Code, Generate the Correct Terms for Your Paper</concept_desc>
  <concept_significance>100</concept_significance>
 </concept>
</ccs2012>
\end{CCSXML}

\ccsdesc[500]{Do Not Use This Code~Generate the Correct Terms for Your Paper}
\ccsdesc[300]{Do Not Use This Code~Generate the Correct Terms for Your Paper}
\ccsdesc{Do Not Use This Code~Generate the Correct Terms for Your Paper}
\ccsdesc[100]{Do Not Use This Code~Generate the Correct Terms for Your Paper}

\end{comment}



\maketitle

\section{Introduction}\label{sec:intro}

Recent advances in AI coding assistants combine the promise of creating
more code faster with the burden of checking that code for correctness.
The most effective approach to checking code correctness at scale
remains testing, and in the case of new bugs, this necessitates
reproduction tests.
A reproduction test is a test that fails on the current code-base to
confirm the presence of an issue and passes in the new code after the
issue has been addressed to confirm its absence.
Reproduction test generation differs from traditional test generation
since it must start from an issue and since the new code on which a
reproduction test should pass does not yet exist at test generation time.

Unfortunately, while there has been good progress towards reproduction test
generation for Python~\cite{ahmed_et_al_2025,khatib_mathews_nagappan_2026,mundler_et_al_2024,wang_et_al_2024},
this is less true for other languages.
For example, Java is widely used in industry thanks to its enterprise
software ecosystem, but reproduction test generation for Java lags
behind that for Python.
Because of Java's static type system and object-oriented nature,
workflows and models optimized for Python tend to perform less well
for Java~\cite{rashid_et_al_2025,zan_et_al_2025}.
This is exacerbated by the fact that bugs in enterprise Java
applications do not follow the same distribution as bugs in
open-source Python repositories.
On the positive side, while Python benchmarks are struggling with
saturation and overfitting~\cite{liang_garg_moghaddam_2025}, this
problem is less severe with Java benchmarks, making them a more genuine
yardstick for innovation.

Benchmarks and solutions for Java reproduction testing are few.
Libro~\cite{kang_yoon_yoo_2023} is an early reproduction test generator for Java
but has been outperformed by later approaches~\cite{mundler_et_al_2024}.
BRT Agent~\cite{cheng_et_al_2025} was evaluated on 80 bugs across
7~languages including some Java bugs, but that evaluation data is
small and not publicly available.
Both Libro and BRT Agent have an important architectural omission:
they lack fault localization and thus only solve a simplified variant
of the reproduction test generation problem.
The OmniCode~\cite{sonwane_et_al_2026} benchmark includes
Java instances and a test-generation task, but does not evaluate
whether tests reproduce issues on existing (pre-patch) code.
Furthermore, being public, the datasets for Libro and OmniCode are
susceptible to contamination, raising the question how evaluated
solutions perform on industry datasets.

To address these gaps, this paper evaluates reproduction test
generation for Java both on a public dataset (to facilitate
comparisons) and on a proprietary dataset (to avoid contamination and
explore industry relevance).
Our public dataset has 250 instances, large enough for meaningful
results; we have released it as a benchmark.
Our proprietary dataset has 150 instances; while we cannot make it publicly
available, we share insights useful for the research community.
Furthermore, we report our experiences adapting a state-of-the-art
Python reproduction test generator~\cite{ahmed_et_al_2026} to
Java, which involved various changes to improve its effectiveness.
By doing so, we end up with the (to our knowledge) strongest solution
for this task to-date.

The starting point for our Java reproduction test generator is e-Otter++,
an LLM-based workflow~\cite{ahmed_et_al_2026}.
Unlike recent coding agents that give the AI unfettered access to
terminal commands~\cite{wang_et_al_2024,wang_et_al_2025_openhands},
e-Otter++ carefully restricts available actions.
This prevents dangerous side-effects that would be too risky for an
industry setting~\cite{ganhotra_et_al_2026,mirchev_et_al_2026}.
Our solution starts from localizing which code to test and where to
put the new test, followed by generating a test, and using a
feedback-driven refinement loop to improve it.
Furthermore, we use inference scaling to generate multiple
reproduction tests candidates, and finally select one test to submit.
Unlike most work on inference scaling that increases candidate diversity using
the temperature hyperparameter~\cite{ehrlich_et_al_2025,xia_et_al_2025}, our
approach uses heterogeneous prompting~\cite{ahmed_et_al_2026} instead.

%\begin{alltt}\textcolor{red}{TODO}\scriptsize
%%- very short summary of results, with concrete numbers
%\end{alltt}

This paper makes the following contributions:

\begin{enumerate}
  \item TDD-Bench-Java\footnote{\url{https://github.com/IBM/TDD-Bench-Verified/tree/main/TDD-Bench-Java}} is a new benchmark for generating reproduction tests given an issue and a source code repository. 
  We have already open-sourced the benchmark.
  \item e-Otter++ for Java, a strong reproduction-test generator, adapted from
    earlier work on the same task for Python. e-Otter++ achieves 43.6\% and 46.4\% fail-to-pass rate with Claude-Sonnet-4.5 and GPT-5.2, respectively.
  \item Empirical insights from experiments on both the public
    TDD-Bench-Java and a proprietary industry dataset.
\end{enumerate}

Overall, we hope that by introducing TDD-Bench-Java, this paper spurs
progress towards better and more reliable AI-based coding assistance
for Java.

\section{Problem Statement and Background}

This section discusses the problem statement and briefly introduces prior work we build upon:
e-Otter~\cite{ahmed_et_al_2026}, 
SWT-bench~\cite{mundler_et_al_2024}, and TDD-Bench-Verified~\cite{ahmed_et_al_2024}. These works focus on Python rather than Java.

\subsection{Problem Statement}

\begin{figure}[t]
  \centerline{\includegraphics[width=\columnwidth]{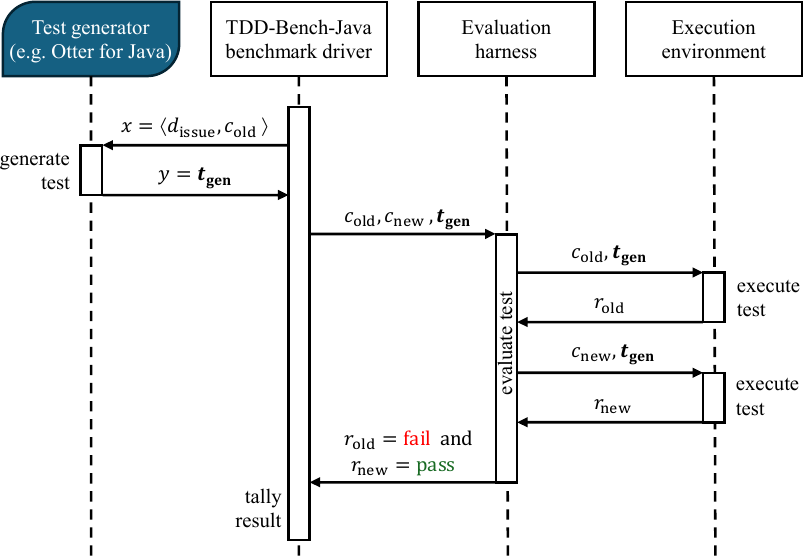}}
  \caption{\label{fig:harness} Evaluation harness for bug reproduction test. 
    First, run the test generator to obtain a test~$t_\textrm{gen}$.
    Second, execute test $t_\textrm{gen}$ on the old code $c_\textrm{old}$
    to obtain an execution result~$r_\textrm{old}$.
    Third, execute test $t_\textrm{gen}$ on the new code $c_\textrm{new}$
    to obtain an execution result~$r_\textrm{new}$.
    Finally, tally whether the test was fail-to-pass, which means
    $r_\textrm{old}$ is fail and $r_\textrm{new}$ is pass.}
\end{figure}

This paper tackles the problem of generating reproduction tests
directly from issue descriptions.
\cref{fig:harness} illustrates the evaluation harness.
The input $x$ to repository-level reproduction test generation
comprises the issue description $d_\textrm{issue}$ and the original
code~$c_\textrm{old}$ on which the issue was reported.
The issue description is typically written in natural language and may
include code snippets or stack traces.
A reproduction test generator, such as Otter for Java, must generate
as output~$y$ a test~$t_\textrm{gen}$ that the benchmark driver then
evaluates.
Crucially, the driver does not reveal the new code~$c_\textrm{new}$
that resolves the issue to the test generator, reflecting real-world
scenarios where fixes are not yet available.
The test generator must work without access to~$c_\textrm{new}$, which
is only used in the evaluation harness.
The evaluation harness executes the generated test~$t_\textrm{gen}$
twice, once each on the code before~($c_\textrm{old}$) and
after~($c_\textrm{new}$) issue resolution.
To get credit for generating a good reproduction test, that test must
be fail-to-pass.
The TDD-Bench-Java benchmark contains 250~instances, each of which has
its own issue description~$d_\textrm{issue}$ and
code~$c_\textrm{old},c_\textrm{new}$.
Repeating the above procedure for all of them enables the benchmark
driver to calculate a fail-to-pass rate.
Formally, the goal is to design a function genTests that takes
\mbox{$x=\langle d_\textrm{issue},c_\textrm{old}\rangle$} as input and
outputs the generated test \mbox{$y=t_\textrm{gen}$}.

\subsection{Reproduction Test Generation Benchmarks}
SWT-bench~\cite{mundler_et_al_2024} and TDD-Bench-Verified~\cite{ahmed_et_al_2024} are both benchmarks designed to evaluate the ability of models 
to generate reproduction tests for software engineering issues, and both are derived from SWE-bench datasets. 
They assess whether a generated test correctly captures an issue by checking 
the fail-to-pass (\failtopass) behavior---i.e., the test should fail on the original code and pass after applying the developer's fix. 
These benchmarks provide standardized evaluation harnesses consistent
with \cref{fig:harness} that simulate real-world development scenarios
where only the issue description~$d_\textrm{issue}$ and pre-fix
code~$c_\textrm{old}$ are available to the test generator.

While both benchmarks share the same goal and evaluation principle, 
they differ slightly in implementation details and dataset construction. 
TDD-Bench-Verified focuses on high-quality, filtered instances where contributing 
tests are explicitly validated, whereas SWT-bench includes broader settings 
and variations such as Lite and Verified subsets, 
and may execute entire test files instead of only contributing tests. 
Despite these differences, prior work shows that model performance is generally consistent across both benchmarks, 
making them complementary tools for evaluating reproduction test generation systems.
Both benchmarks support evaluation for Python, not Java.

\begin{table*}[!t]
\centering
\caption{TDD-Bench-Java benchmark dataset attributes.}
\label{tab:bench}
%\vskip 0.05in
\resizebox{.85\textwidth}{!}{%
\renewcommand{\arraystretch}{1.2}% Tighter
\begin{tabular}{lrrlrrr}
\toprule    
\multicolumn{1}{c}{\multirow{2}{*}{Project}} & \multicolumn{1}{c}{\multirow{2}{*}{\# of Instances}} & \multicolumn{1}{c}{\multirow{2}{*}{\begin{tabular}[c]{@{}c@{}}Fraction of \\ Dataset (in \%)\end{tabular}}} & \multicolumn{1}{c}{\multirow{2}{*}{Build Tools}} & \multicolumn{2}{c}{\begin{tabular}[c]{@{}c@{}}Average \# of Line \\ Deleted and Added\end{tabular}} & \multicolumn{1}{c}{\multirow{2}{*}{\begin{tabular}[c]{@{}c@{}}Average Word Count\\ in Issue Description\end{tabular}}} \\ \cline{5-6}
 \multicolumn{1}{c}{}                         & \multicolumn{1}{c}{} & \multicolumn{1}{c}{} & \multicolumn{1}{c}{} & \multicolumn{1}{c}{on Code} & \multicolumn{1}{c}{On Tests} & \multicolumn{1}{c}{} \\ \midrule
 trinodb/trino                                & 44                   & 17.6                 & Maven                & 101.2                       & 82.8                         &  98.5                \\
 fasterxml/jackson-databind                   & 42                   & 16.8                 & Maven                &  39.6                       & 63.2                         & 198.2                \\
 apache/rocketmq                              & 41                   & 16.4                 & Maven                &  61.1                       & 82.8                         & 190.6                \\
 apache/dubbo                                 & 36                   & 14.4                 & Maven                &  59.8                       & 159.8                        & 181.2                \\
 google/gson                                  & 34                   & 13.6                 & Maven                &  67.5                       & 108.6                        & 198.0                \\
 fasterxml/jackson-core                       & 18                   &  7.2                 & Maven                &  38.7                       & 68.0                         & 115.7                \\
 alibaba/fastjson2                            &  6                   &  2.4                 & Maven                &  12.8                       & 31.8                         &  89.2                \\
 apolloconfig/apollo                          &  6                   &  2.4                 & Maven                & 117.8                       & 62.8                         & 113.0                \\
 mockito/mockito                              &  6                   &  2.4                 & Gradle               & 101.8                       & 99.5                         & 160.7                \\
 elastic/logstash                             &  5                   &  2.0                 & Gradle               &  64.0                       & 39.8                         & 294.8                \\
 fasterxml/jackson-dataformat-xml             &  5                   &  2.0                 & Maven                & 105.0                       & 65.8                         & 200.8                \\
 googlecontainertools/jib                     &  5                   &  2.0                 & Gradle               &  21.4                       & 26.2                         & 235.0                \\
 google/guava                                 &  2                   &  0.8                 & Maven                &  59.0                       & 92.5                         &  94.5                \\\midrule
 Average	& 19.5	&7.8	&NA	&86.3	&87.2	&199.4 \\ \bottomrule
\end{tabular}
}
\end{table*}

\subsection{Reproduction Test Generation for Python}
Reproduction test generation plays a critical role in resolving software engineering issues. 
Prior approaches either rely on zero-shot prompting with LLMs or incorporate existing tests as contextual 
input~\cite{mundler_et_al_2024, ahmed_et_al_2024}. 
However, recent studies have demonstrated that zero-shot methods underperform compared to 
approaches (e.g., Otter~\cite{ahmed_et_al_2025}, e-Otter~\cite{ahmed_et_al_2026}, AEGIS~\cite{wang_et_al_2024}, 
AssertFlip~\cite{khatib_mathews_nagappan_2026}) that leverage relevant portions of the codebase.
Our prior work, Otter, is an LLM-based system that combines localization, a self-reflective action planner, and test generation. 
It identifies relevant files and functions, plans actions (read/write/modify), 
and generates tests while using rule-based validation and repair.

e-Otter is an improved version of Otter, where we apply execution 
feedback on the old codebase ($c_\textrm{old}$) along with an LLM-based critic to refine the tests. 
We also increase test candidate diversity through issue morphing, where an LLM rewrites 
the issue description. Experiments show that both execution feedback and issue morphing significantly 
improve performance. 
Finally, we select the best test based on execution feedback from candidate test patches, which we refer to as e-Otter++.
In this paper, we modify our pipeline to make it work for Java, and we discuss the approach in~\cref{sec:method}.

\section{TDD-Bench Java Benchmark}\label{sec:benchmark}

Resolving an issue involves modifying code to create a patch that fixes 
the problem and is commonly used to evaluate LLM-based coding agents. 
However, existing benchmarks such as SWE-bench~\cite{jimenez_et_al_2024} are largely 
limited to Python, making them insufficient for assessing model 
performance across diverse software ecosystems. 
To address this limitation, recent work introduces two complementary multi-language benchmarks: 
Multi-SWE-bench~\cite{zan_et_al_2025} and SWE-PolyBench~\cite{rashid_et_al_2025}. 
Multi-SWE-bench expands evaluation to seven programming languages (Java, TypeScript, JavaScript, Go, Rust, C, and C++), 
comprising 1,632 carefully curated and human-validated instances.
%, and is accompanied by a large-scale RL dataset (Multi-SWE-RL) to support future training. 
In parallel, SWE-PolyBench provides 2,110 
instances across four languages (Python, Java, JavaScript, and TypeScript), covering a broader range of tasks, including bug fixes, 
feature additions, and refactoring.
It also introduces novel evaluation metrics based on syntax tree analysis to better capture code 
understanding and localization capabilities.
Both are repository-level, execution-based benchmark.
Together, they enable comprehensive evaluation of state-of-the-art coding agents across languages, 
task types, and complexity levels. 
However, neither benchmark is designed to evaluate the quality of reproduction tests. 
To address this gap, we use Java samples from both benchmarks and propose a new benchmark for evaluating reproduction tests.

\subsection{Data Filtering}
There are 128 and 165 Java samples in Multi-SWE-bench and SWE-PolyBench, respectively. 
We start with these samples and attempt to reproduce the issues using golden developer-written tests. 
We find that, for some instances, the tests pass on both $c_\textrm{old}$ and $c_\textrm{new}$, and for others, 
the contributing tests (updated or modified in the golden test patch) do not exhibit fail-to-pass behavior. 
There are two instances that appear in both benchmarks. 
After filtering for fail-to-pass behavior and deduplication,
we end up with 250 instances. 
\cref{tab:bench} presents the attributes of the benchmark instances. 
In SWE-PolyBench, all projects use Maven as the build tool, whereas in Multi-SWE-bench, three projects use Gradle. 

\subsection{Evaluation Harness and Metric}
\cref{fig:harness} illustrates how the evaluation harness works in this setup. 
We first execute the reproduction test on $c_\textrm{old}$, where the test should fail to reproduce the issue before the fix. 
After applying the developer- written code patch, we execute the test again; this time, it should pass, 
confirming that the issue has been resolved. 
Both SWE- PolyBench and Multi-SWE-bench provide pre-built Docker environments for test execution. 
However, they are configured to run the entire test suite rather than a specific test. 
We primarily modify the SWE-PolyBench framework to support evaluation of reproduction tests. 
We design our harness to be as flexible as possible, allowing multiple ways to execute reproduction tests. 
Users can provide a class-function name pair or just a class name. 
The harness can also automatically identify the test file and function from test patch if none are specified. 
%We share our benchmark in the supplementary material and plan to open-source it soon.

We use the fail-to-pass rate as the primary evaluation metric, following prior work~\cite{mundler_et_al_2024, ahmed_et_al_2024}. 
The fail-to-pass rate is the percentage of generated tests that fail on the original code ($c_\textrm{old}$)
 and pass after applying the fix ($c_\textrm{new}$), indicating successful reproduction and validation of the issue. 
 SWT-Bench also hosts a leaderboard\footnote{\url{https://swtbench.com/?results=verified}} for Python reproduction test generators, 
 where ranking is based on the fail-to-pass rate. 
 One key difference between Java and Python is that Java requires compilation and build steps, unlike Python. 
 If a Java test attempts to access a class or function that has not yet been defined or will be introduced in the code patch, 
 it results in a build error instead of a test failure. 
 In our benchmark, we treat such build errors as equivalent to test failures,
 whereas in Python, similar issues would typically result in runtime errors.

\section{Methodology}\label{sec:method}

\begin{figure}
  \centerline{\includegraphics[width=\columnwidth]{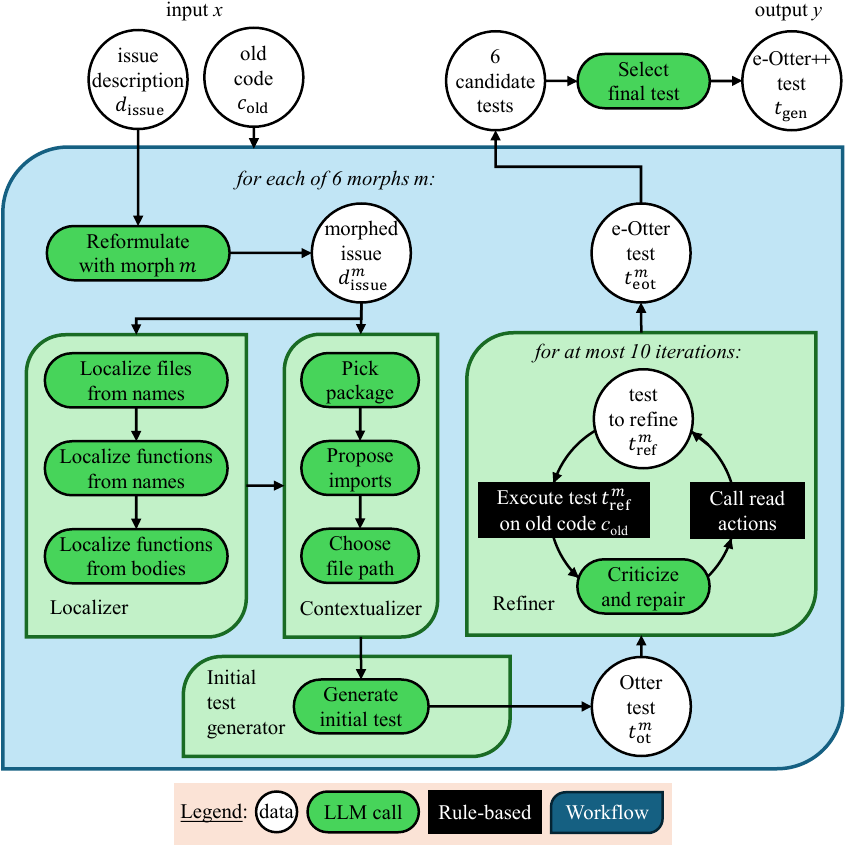}}
  \caption{\label{fig:overview}Overview of test generation pipeline.}
  \vspace{-5pt}
\end{figure}

\cref{fig:overview} illustrates our reproduction test generation
pipeline.
The pipeline uses an execution-free test generation approach~(Otter)
to generate the initial test, then uses execution feedback to refine
the test~(e-Otter), and finally uses inference scaling to further
increase test quality~(e-Otter++).
In other words, the Otter and e-Otter tests can be viewed as ablations
of the final e-Otter++ test; they are named after and inspired by our
prior work with Python~\cite{ahmed_et_al_2025,ahmed_et_al_2026}.
Due to the properties of the TDD-Bench-Java dataset, model advances,
and insights from our proprietary data, we made some modifications to
the Python pipeline, discussed below.

\subsection{Test Generator}

This section discusses the core of our test generation pipeline,
which has four components: localizer, contextualizer,
initial test generator, and refiner.

\vspace{.2cm}

\noindent \textbf{Localizer:}
To generate a fail-to-pass reproduction test, it helps to first
localize \emph{focal} functions, which are the functions under test
that will likely be updated as part of fixing the issue.
In addition, it also helps to localize related regression tests
functions, which usually do not reproduce the issue but cover the
focal functions.
The localizer makes three LLM calls.
The first call presents the issue description and the names of all
Java files in the repository to the LLM, and the LLM selects 50
relevant files from the list~(prioritizing recall over precision).
In most cases, the issue description contains enough information to
help LLMs select the files where the changes will be made.
The second call presents those file names along with the names of
functions defined in them and asks the LLM to choose the relevant
functions from the list.
This yields a set of function-file pairs that can be used as context
for test generation.
So far, the LLM has not seen any function bodies to the LLM, making
its selections based solely on the file and function names.
The final call presents the function body and asks the LLM to filter
out unnecessary functions.
This final filtered set of file-function pairs serves as input for the
contextualizer.

Our original Otter work for Python~\cite{ahmed_et_al_2025} had
two separate localizations for focals and tests, whereas for Java,
we merge them.
This change was motivated by dataset characteristics.
In TDD-Bench-Verified for Python, new test files were written for less
than 1\% of instances.
Instead, tests were primarily written or modified in existing test files.
Localizing tests is very important in that setup, and we had a separate test localizer to identify relevant tests. 
In Java, about 15\% of instances involve tests written in new files in open-source projects. 
In our proprietary data, we have observed that the majority of tests are written in new files, and for some instances, 
there are fewer than five existing test files. Reading irrelevant tests may hurt the performance of the test generator.
By providing a single merged localization step, we effectively leave
it up to the LLM to skip test localization if it determines that no
existing tests are relevant.

\vspace{.2cm}

\noindent \textbf{Contextualizer:} As discussed earlier, in Java, new test files are frequently written, and relevant test files are missing for many instances. 
Unlike Python, writing a test in a file with an incompatible package can cause build errors. 
To address this challenge and avoid irrelevant test context, we decide to write one test file for each instance. 
A Java test file has different components (e.g., package name and relevant imports).
Therefore, the contextualizer starts with the localized functions from
the localizer step and tries to construct these components before
writing the full test file.

For the package name, we gather all existing package statements from the repository and then ask the LLM to choose from them, 
given the issue description and localized functions. 
We frame this as a selection problem to avoid hallucination by the model. 
However, the model will have the opportunity to fix an incorrect package based on execution feedback in the refiner step. 
Similar to the package name, we also determine the imports necessary for the test file. 
We collect all imports from the initially selected 50 files and ask the LLM to choose the required imports from them to provide better context for writing the test.

Another challenge in writing a new test file is determining its location in the repository. 
To address this, we collect all paths to existing directories under the test directory and, similar to package selection, 
ask the LLM to choose the appropriate location for the new test file. 
All in all, the contextualizer makes three LLM calls, and each call
uses the output from the previous call as additional context.

\vspace{.2cm}

\noindent \textbf{Initial Test Generator:} The initial test
generator generates the Otter test~$t_\textrm{ot}^m$ as a single new
function within a new class in a new file.  It does so via a single
LLM call whose input is the issue description and the information
gathered by the preceding localizer and contextualizer steps, prompted
to generate a fail-to-pass reproduction test.  As discussed above,
generating a new file is easier for the LLM and consistent with the
characteristics of our proprietary dataset.  We ask the LLM to
generate a test class name that ends with ``OtterTest'' to avoid name
conflicts and clearly mark which tests were added by the LLM.

%However, writing a new
%test file introduces one issue: it may propose a test file with a name
%that already exists.  To address this conflict, we ask the LLM to
%generate a test class name that ends with ``OtterTest''.

\vspace{.2cm}

\noindent \textbf{Refiner:}
The refiner starts by setting the Otter test~$t_\textrm{ot}^m$ from
the previous step as the initial test to refine~$t_\textrm{ref}^m$.
It iteratively refines it using execution feedback.
To get that execution feedback, it executes test~$t_\textrm{ref}^m$ on
the current code~$c_\textrm{old}$ in the repository.
Next, it presents the LLMs with the issue description and execution
logs on~$c_\textrm{old}$, and asks whether the test is failing for the
right reason: the problem described in the issue description.
If the LLM-based critic determines that the test is failing for the
right reason, the loop immediately terminates.
Otherwise, the LLM rewrites the test, notes the changes made, and
requests any function or class information the model needs; this all
happens in the same single LLM call that also implements the critic.
The refiner gathers this additional information for future attempts
when the critic determines that the test is still not failing for the
right reason.
The refiner loop iterates at most ten times if the critic is not
satisfied.

We cumulatively collect the changes and present them to the LLM to prevent it from generating same buggy code. 
For better context, we try to retrieve the functions and classes proposed by the LLM. In the class information, we include the package name, imports, and caller–callee functions, 
which may help the LLM perform better.
After the loop is done, the refiner returns the final~$t_\textrm{ref}^m$,
which we refer to as the e-Otter test~$t_\textrm{eot}^m$.

\vspace{.2cm}
We share all the prompts used for test generation in the supplementary material\footnote{\url{https://drive.google.com/drive/folders/1i6Tp2AvlN7x7lWcGw_npeZIZvGGWleWl?usp=sharing}}.

\subsection{Heterogeneous prompting}
Inference scaling is beneficial for reproduction test
\mbox{generation~\cite{ahmed_et_al_2025,ahmed_et_al_2026}}.
Our prior work demonstrated that mask- and morph-based inference
scaling perform better than traditional temperature-based
scaling~\cite{ahmed_et_al_2026}.
It also demonstrated that morphing yielded relatively better results
than masking.
However, with newer advanced models, we did not observe any benefit
from masking on the Java dataset.
After some initial experiments, we decided to move forward with
morphing only.
In morph-based inference scaling, the LLM rewrites the issue
description to increase the diversity of test candidates.
Our previous work contains an extensive discussion of morphing
including relevant prompts~\cite{ahmed_et_al_2026}. 
For completeness, we briefly discuss them below.
As shown in \cref{fig:overview}, each morph~$m$ yields one rewritten
issue description~$d_\textrm{issue}^m$.

\vspace{.2cm}

\noindent \textbf{standard:} This morphing technique asks the LLM
to rewrite the issue description in a standard format.
A well-written issue description includes several components, 
such as a title, description, steps to reproduce, and expected behavior. 
These components may help improve the generation of reproduction tests.

\vspace{.2cm}

\noindent \textbf{simple:} Some issue descriptions contain code artifacts and use project-specific jargon, which can make 
it difficult to understand the problem. 
This morph asks the LLM to simplify the issue description 
so that it becomes easier for the model to perform better.

\vspace{.2cm}

\noindent \textbf{dropCode:} Some issue descriptions contain misleading or irrelevant code snippets that can confuse the model. 
For example, an issue may include code using a library that is not actually part of the project, 
leading the model to generate incorrect tests. 
Removing such code (dropCode morph) helps the model focus on the core problem and avoid making invalid assumptions.

\vspace{.2cm}

\noindent \textbf{initTest:} In the initTest morph, the LLM is asked to generate an initial fail-to-pass test and incorporate it into the issue description without additional context. 
This provides a starting point for the test generation process, allowing subsequent stages to refine and improve the test.

\vspace{.2cm}

\noindent \textbf{initPatch:} In the initPatch morph, the LLM is prompted to propose an initial solution or code patch for the issue and include it in the description. 
This helps guide the model by providing insight into the expected behavior or potential fix, which can improve the quality of generated tests.

\vspace{.2cm}

\noindent \textbf{default:} Our test generation pipeline and
experiments also includes one setting with the original issue
description, without any LLM-based issue rewriting.

%\vspace{-.225cm}

\subsection{Test Selector}
In real development settings, developers typically prefer a single test.
Therefore, after inference scaling, we need to select the best test
from a pool of candidates.
To address this, the test selector prompts an LLM with six test
patches~$t_\textrm{eot}^m$ for the six morphs~$m$, along with their
execution logs on~$c_\textrm{old}$.
The LLM then selects the best test from the pool based on these logs. 
We instruct the LLM to critically analyze and compare the tests before choosing one. The prompt is provided in the supplementary material.
We refer to this single final generated test~$t_\textrm{gen}$ as the
e-Otter++ test.

\section{Results on TDD-Bench-Java}

This section reports results on our new public benchmark
TDD-Bench-Java~(see \Cref{sec:benchmark}) using our approach
e-Otter++~(see \Cref{sec:method}).
It also drills down into the impact of key components of our approach,
execution feedback and heterogeneous prompting.

\subsection{Experimental Setup}
We used two models, Claude-Sonnet-4.5 and GPT-5.2, for our experiments. 
The number of LLM calls is evident from \Cref{fig:overview}:
seven to generate each initial test~$t_\textrm{ot}^m$ plus
up to~10 for refinement to generate each test~$t_\textrm{eot}^m$.
Inference scaling adds one more call per non-default morph,
and there is a final selection call to choose~$t_\textrm{gen}$.

\subsection{Performance of Test Generators}

\begin{figure}
  \centerline{\includegraphics[width=\columnwidth]{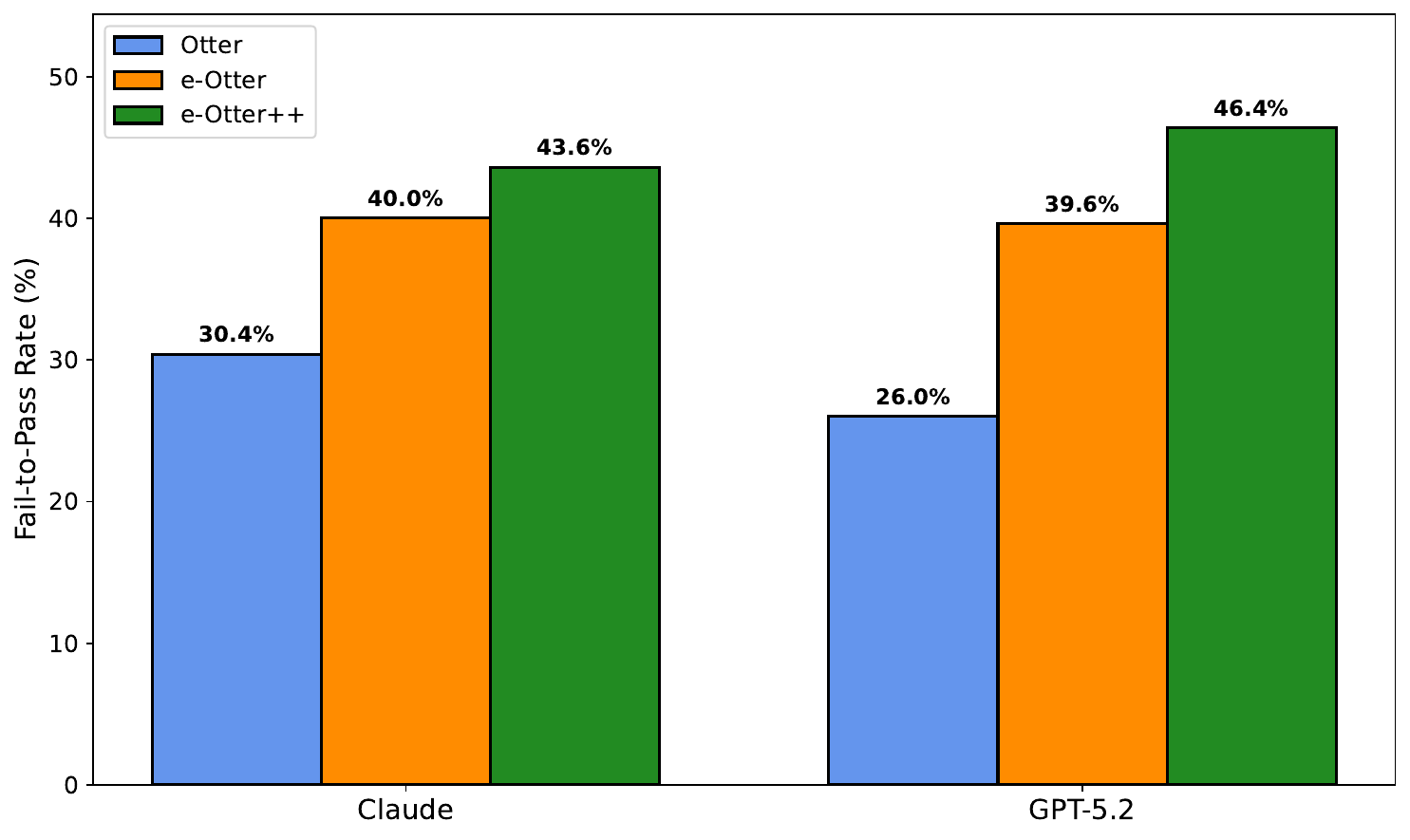}}
  \caption{\label{fig:result} Performance of Otter, e-Otter, and e-Otter++ on TDD-Bench Java.}
\end{figure}

\cref{fig:result} presents the performance of Otter, e-Otter, and e-Otter++ using two models: Claude-Sonnet-4.5 and GPT-5.2. 
We observe that Claude-Sonnet-4.5 performs better than GPT-5.2 with Otter and e-Otter, while GPT-5.2 outperforms Claude with e-Otter++. 
The improvement from refinement is higher for GPT-5.2 (13.6 points vs. 9.4 points). To measure statistical significance, we perform McNemar’s test~\cite{mcnemar1947note}.
McNemar’s test is applicable to this setup because it evaluates statistical significance on paired binary outcomes (fail-to-pass vs. not), 
which aligns with comparing model performance on the same set of instances. 
We observe statistical significance ($p < 0.01$) for both models when comparing Otter and e-Otter. 
For e-Otter and e-Otter++, we only achieve statistical significance
for GPT-5.2, but not for the Claude model.

\vspace{.1cm}
\findings{1}{We observe a 43.6\% and 46.4\% fail-to-pass rate with Claude-Sonnet-4.5 and GPT-5.2, respectively, using e-Otter++.}

\subsection{Impact of Refiner}

\cref{tab:refiner} presents the results for both Otter and e-Otter across all six test candidates based on different morphs~$m$. 
The performance of e-Otter is similar for both Claude-Sonnet-4.5 and GPT-5.2. The fail-to-pass rate ranges from 38.8\% to 40.4\% for Claude, 
while for GPT-5.2 it ranges from 36.8\% to 41.6\%. 
However, the relative performance improvement for GPT-5.2 (43.9\%–64.3\%) is higher than that of the Claude model (26.2\%–54.7\%) across all candidates. 
We also performed McNemar’s test to assess statistical significance and observed that the improvement is significant ($p < 0.01$) for all model–candidate pairs.

\vspace{.1cm}
\findings{2}{Using refinement, the relative improvement from Otter to e-Otter ranges from 43.9\%–64.3\% for GPT-5.2 and 26.2\%–54.7\% for Claude across all candidates.}

\begin{table}[t]
\centering
\caption{Performance of Otter and e-Otter with different morphs and models.}
\label{tab:refiner}
%\vskip 0.05in
\resizebox{.95\columnwidth}{!}{%
\renewcommand{\arraystretch}{1.2}% Tighter
\begin{tabular}{llrrrrr}
\toprule    
\multicolumn{1}{c}{\multirow{2}{*}{Model}} & \multicolumn{1}{c}{\multirow{2}{*}{Prompt}} & \multicolumn{2}{c}{Otter}                           & \multicolumn{2}{c}{e-Otter}                         & \multicolumn{1}{c}{\multirow{2}{*}{Change in \%}} \\
\multicolumn{1}{c}{}                       & \multicolumn{1}{c}{}                        & \multicolumn{1}{c}{\failtopass} & \multicolumn{1}{c}{in \%} & \multicolumn{1}{c}{\failtopass} & \multicolumn{1}{c}{in \%} & \multicolumn{1}{c}{}                              \\ \midrule
\multirow{6}{*}{Claude-Sonnet-4.5} & default                 & 76         & 30.4         & 100         & 40.0            & 31.6                         \\
                                   & standard                & 64         & 25.6         & 99          & 39.6          & 54.7                        \\
                                   & simple                  & 71         & 28.4         & 97          & 38.8          & 36.6                         \\
                                   & dropCode                & 68         & 27.2         & 99          & 39.6          & 45.6                         \\
                                   & initTest                & 71         & 28.4         & 99          & 39.6          & 39.4                         \\
                                   & initPatch               & 80         & 32.0           & 101         & 40.4          & 26.2                         \\ \midrule
\multirow{6}{*}{GPT-5.2}           & default                 & 65         & 26.0           & 99          & 39.6          & 52.3                         \\
                                   & standard                & 63         & 25.2         & 102         & 40.8          & 61.9                          \\
                                   & simple                  & 68         & 27.2         & 98          & 39.2          & 44.1                         \\
                                   & dropCode                & 56         & 22.4         & 92          & 36.8          & 64.3                         \\
                                   & initTest                & 66         & 26.4         & 95          & 38.0            & 43.9                         \\
                                   & initPatch               & 67         & 26.8         & 104         & 41.6          & 55.2           \\ \bottomrule                                        
\end{tabular}
}
\end{table}

\begin{table}[t]
\centering
\caption{Impact of heterogeneous prompts. The `All' rows evaluate \failtopass @ 6, whereas rows of the form \mbox{`All - $d$'} evaluate \failtopass @ 5, since they ablate morph~$d$ from the candidate pool.}
\label{tab:hetero}
%\vskip 0.05in
\resizebox{.95\columnwidth}{!}{%
\renewcommand{\arraystretch}{1.2}% Tighter
\begin{tabular}{llrrr}
\toprule    
\multicolumn{1}{c}{\multirow{2}{*}{Model}}                                       & \multicolumn{1}{c}{\multirow{2}{*}{Prompt}} & \multicolumn{1}{c}{\multirow{2}{*}{\failtopass @ N}} & \multicolumn{1}{c}{\multirow{2}{*}{in \%}} & \multicolumn{1}{c}{\multirow{2}{*}{Change in \%}} \\
\multicolumn{1}{c}{}                                                             & \multicolumn{1}{c}{}                        & \multicolumn{1}{c}{}                         & \multicolumn{1}{c}{}                       & \multicolumn{1}{c}{}                              \\ \midrule
\multirow{7}{*}{Claude-Sonnet-4.5} & All                     & 137                      & 54.8                   & NA                            \\
                                   & All - default           & 136                      & 54.4                   & -0.7                          \\
                                   & All - standard          & 134                      & 53.6                   & -2.2                          \\
                                   & All - simple            & 136                      & 54.4                   & -0.7                          \\
                                   & All - dropCode          & 133                      & 53.2                   & -2.9                          \\
                                   & All - initTest          & 135                      & 54.0                     & -1.5                          \\
                                   & All - initPatch         & 136                      & 54.4                   & -0.7                          \\ \midrule
\multirow{7}{*}{GPT-5.2}           & All                     & 137                      & 54.8                   & NA                            \\ 
                                   & All - default           & 133                      & 53.2                   & -2.9                          \\
                                   & All - standard          & 133                      & 53.2                   & -2.9                          \\
                                   & All - simple            & 137                      & 54.8                   & 0.0                             \\
                                   & All - dropCode          & 132                      & 52.8                   & -3.6                          \\
                                   & All - initTest          & 135                      & 54.0                     & -1.5                          \\
                                   & All - initPatch         & 134                      & 53.6                   & -2.2  \\ \bottomrule                                           
\end{tabular}
}
\end{table}

\subsection{Impact of Heterogeneous Prompting}
In our prior work~\cite{ahmed_et_al_2026}, we demonstrated that heterogeneous prompting helps increase test candidate diversity for Python. 
Does this hold for Java as well? 
\cref{tab:refiner} shows that with a single test variant, we can achieve at most 101 and 104 fail-to-pass cases with Claude-4.5-Sonnet and GPT-5.2, respectively. 
However, 
\cref{tab:hetero} shows that the fail-to-pass @N (N=6) reaches 54.8\% (137) for both models.
This indicates that heterogeneous prompting increases the fail-to-pass rate by more than 32\%. 
Fail-to-pass @N is the proportion of instances for which at least one out of N generated test candidates fails on the old code and passes on the new code.  
\cref{tab:hetero} also presents ablation results showing the contribution of each test variant. 
We observe that each variant generates 0–5 fail-to-pass tests that are not resolved by other candidates. 
Although the unique contribution of each variant is relatively small, it still benefits the test selector by providing more fail-to-pass tests to choose from.

\vspace{.1cm}
\findings{3}{Heterogeneous prompting increases the fail-to-pass rate @N by more than 32\% for both Claude-Sonnet-4.5 and GPT-5.2 models.}

\subsection{An Illustrative Example}
\begin{figure}
  \raggedright
%%%%%%%%%%%%%%%%%%%%
  \underline{issue description $d_\textrm{issue}$}\\
  \begin{lstlisting}
Inconsistent TextBuffer#getTextBuffer behavior
Hi, I'm using 2.4.2. While I'm working on CBORParser, 
I noticed that CBORParser#getTextCharacters() returns
 sometimes `null` sometimes `[]` (empty array) when 
 it's parsing empty string `""`.

While debugging, I noticed that TextBuffer#getTextBuffer
 behaves inconsistently.

```
TextBuffer buffer = new TextBuffer(new BufferRecycler());
buffer.resetWithEmpty();
buffer.getTextBuffer(); // returns null
buffer.contentsAsString(); // returns empty string ""
buffer.getTextBuffer(); // returns empty array []
```

I think getTextBuffer should return the same value. 
Not sure which (`null` or `[]`) is expected though.
  \end{lstlisting}

%%%%%%%%%%%%%%%%%%%%  
  \underline{Developer Written Test}
  \begin{lstlisting}[language=python]
public void testEmpty() {
  TextBuffer tb = new TextBuffer(new BufferRecycler());
  tb.resetWithEmpty();
  assertTrue(tb.getTextBuffer().length == 0);
  tb.contentsAsString();
  assertTrue(tb.getTextBuffer().length == 0);
}
 \end{lstlisting}

%%%%%%%%%%%%%%%%%%%%
  \underline{Otter Generated Test}
  \begin{lstlisting}[language=python]
package com.fasterxml.jackson.core.util;

import com.fasterxml.jackson.core.util.BufferRecycler;
import com.fasterxml.jackson.core.util.TextBuffer;
import static org.junit.Assert.*;
import org.junit.Test;

public class TextBufferConsistencyOtterTest {
    
    @Test
    public void testGetTextBufferConsistency() {
        TextBuffer buffer = new TextBuffer(new BufferRecycler());
        buffer.resetWithEmpty();
        
        // First call to getTextBuffer
        char[] firstResult = buffer.getTextBuffer();
        
        // Call contentsAsString which may modify internal state
        String str = buffer.contentsAsString();
        assertEquals("", str);
        
        // Second call to getTextBuffer should return the same 
        // as first call
        char[] secondResult = buffer.getTextBuffer();
        
        // Both should be non-null and consistent
        assertNotNull("First getTextBuffer() 
              call should not return null", firstResult);
        assertNotNull("Second getTextBuffer() 
              call should not return null", secondResult);
        assertSame("getTextBuffer() should return the 
              same reference on consecutive calls", 
                   firstResult, secondResult);
    }
}
  \end{lstlisting}
%%%%%%%%%%%%%%%%%%%%
  \vspace*{-2mm}
  \caption{\label{fig:example}e-Otter++ generated test forfasterxml\_\_jackson-core-183.}
  \vspace*{-2mm}
\end{figure}

\noindent\cref{fig:example} shows an example of an e-Otter++ generated test for ``forfasterxml\_\_jackson-core-183''. 
We also present the developer-written test. 
Both tests reproduce the issue correctly but differ significantly in nature. 
The issue asks for consistent behavior. 
The developer test ensures this by comparing lengths, whereas the Otter-generated test expects the same object instance, 
which is too strict compared to the developer-written test. 
Although the two tests are qualitatively different, they both successfully reproduce the bug. 
A deeper qualitative comparison between model-generated tests and developer-written tests is 
beyond the scope of this paper and is left for future work.

\section{Reproduction Test Generation on Proprietary Projects}

One of the challenges of applying LLMs to open-source data is model 
contamination. The model may have already seen the data, which may 
impact the overall findings. Some benchmarks address this problem by 
filtering issues that appear after a certain model cutoff date~\cite{badertdinov_et_al_2025}. 
However, even for those benchmarks, 
the evaluation may not be bias-free compared to proprietary data, 
since the model may have seen prior versions of the open-source 
project and acquired background knowledge of the repository that 
can help it perform better. 
At the same time, businesses seek to benefit from LLMs on their own
proprietary code.
To this end, this section shares our findings on industry code.

\begin{figure*}[htbp]
    \centering
    \begin{subfigure}[b]{0.24\textwidth}
        \centering
        \includegraphics[width=\textwidth]{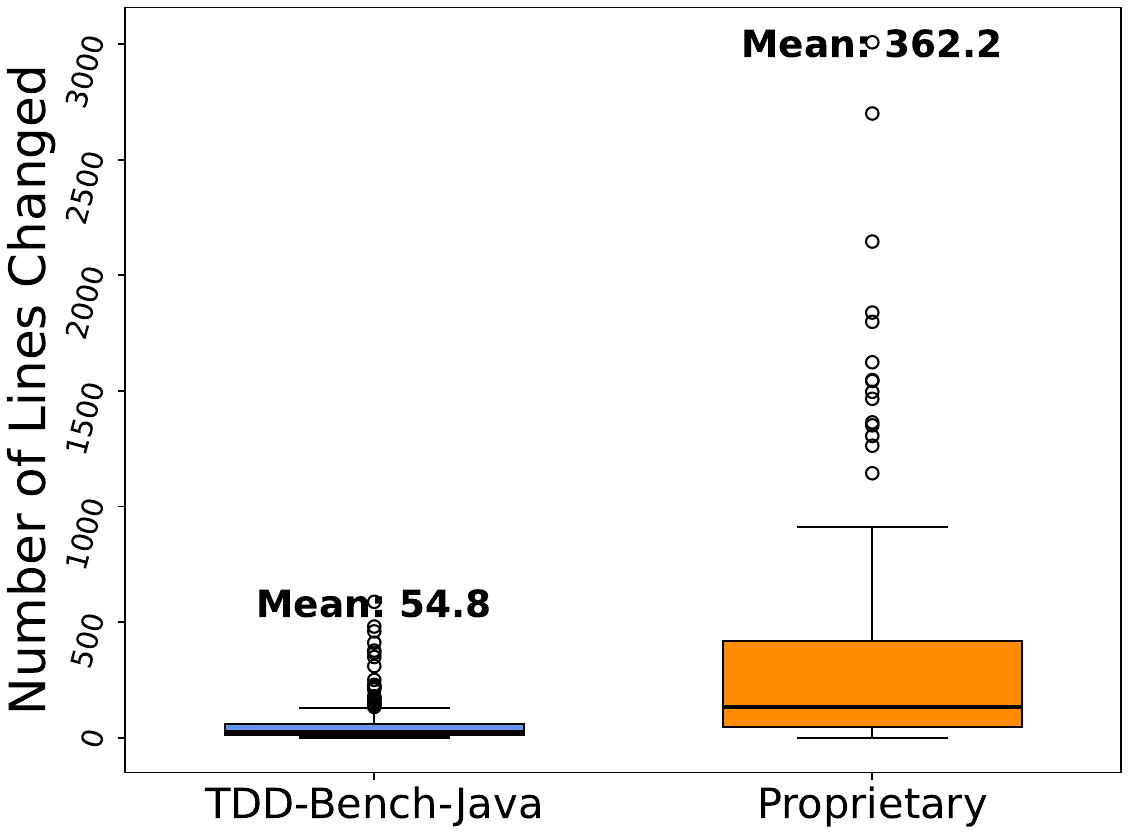}
        \caption{Number of changed lines}
        \label{fig:1}
    \end{subfigure}
    \hfill
        \begin{subfigure}[b]{0.24\textwidth}
        \centering
        \includegraphics[width=\textwidth]{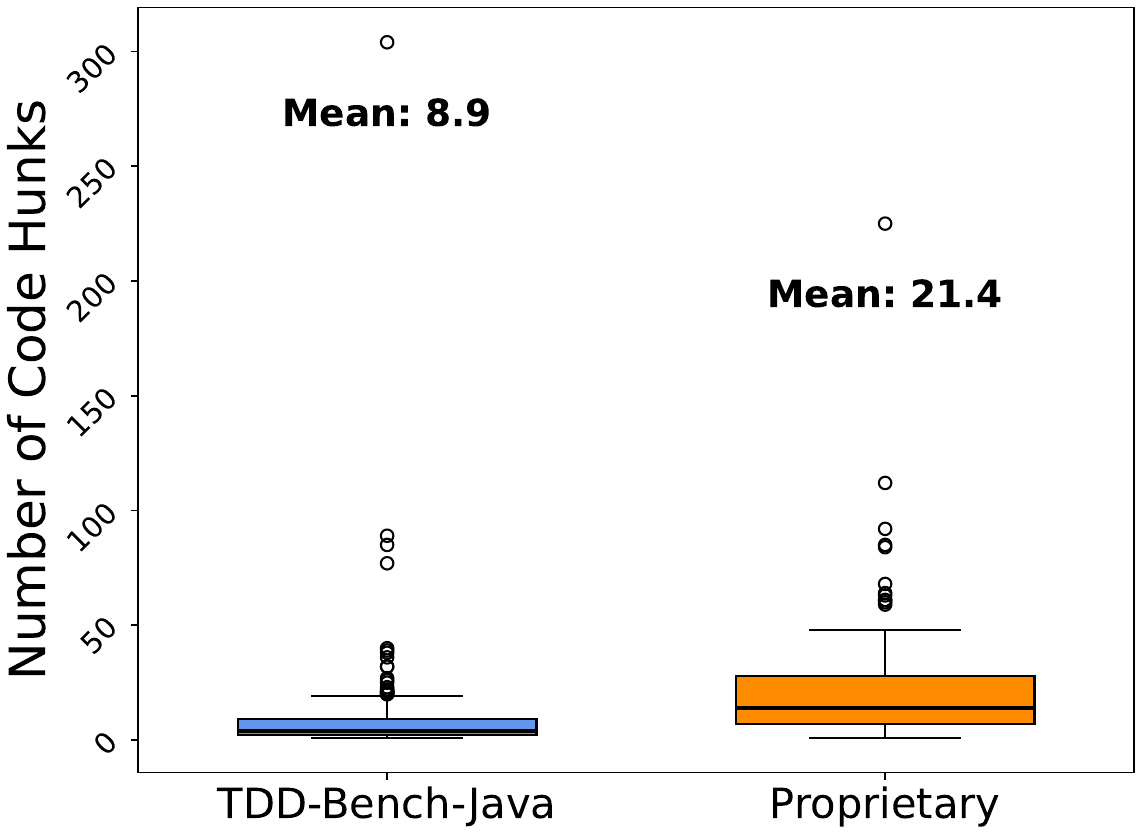}
        \caption{Number of diff-hunks}
        \label{fig:4}
    \end{subfigure}
    \hfill
    \begin{subfigure}[b]{0.24\textwidth}
        \centering
        \includegraphics[width=\textwidth]{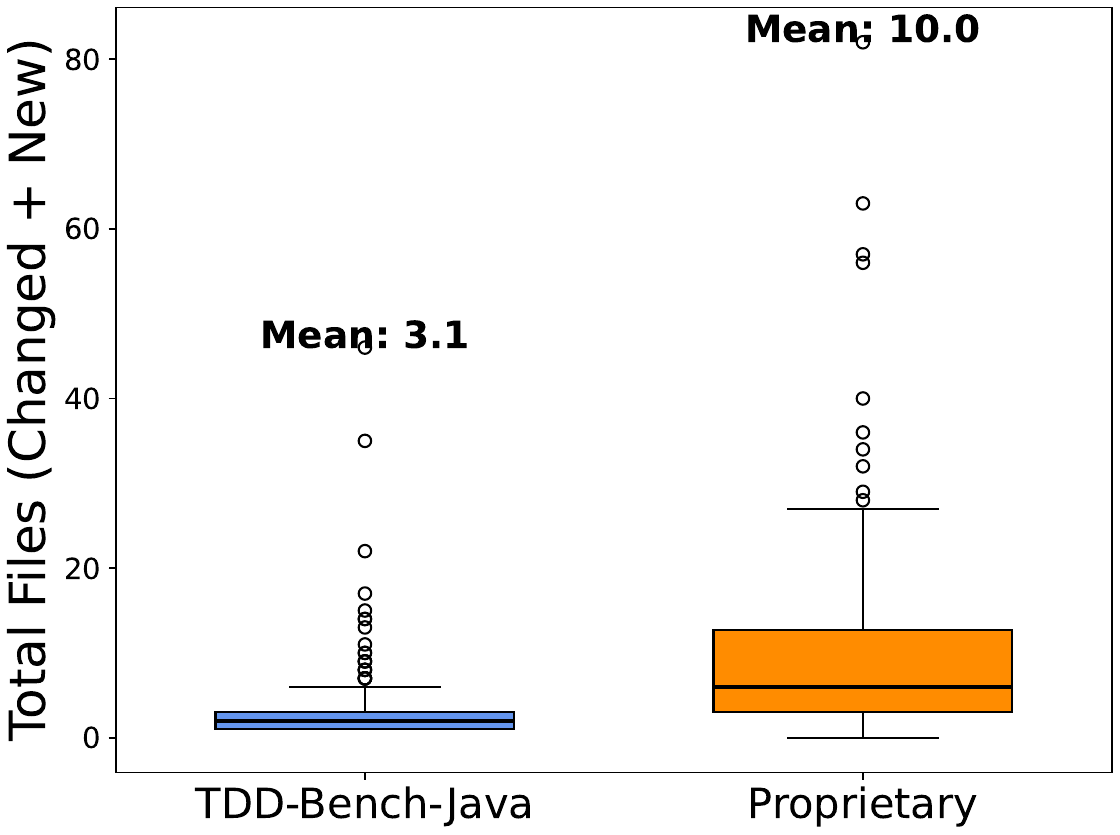}
        \caption{Total files (modified + new)}
        \label{fig:2}
    \end{subfigure}
    \hfill
    \begin{subfigure}[b]{0.24\textwidth}
        \centering
        \includegraphics[width=\textwidth]{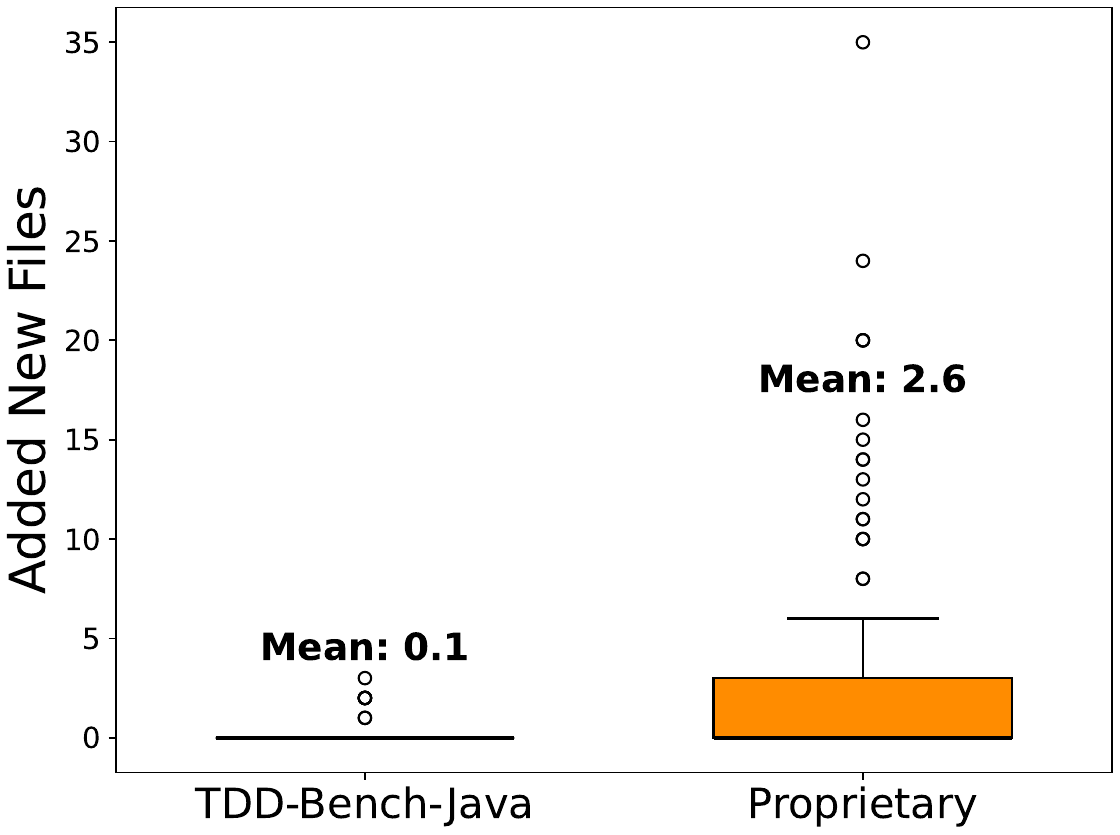}
        \caption{Added new files}
        \label{fig:3}
    \end{subfigure}
    \caption{Comparing the code patches in open-source and proprietary data.}
    \label{fig:three_side}
\end{figure*}

\subsection{Characteristics of Proprietary Data}
\label{Characteristics}
\begin{figure}
  \centerline{\includegraphics[width=\columnwidth]{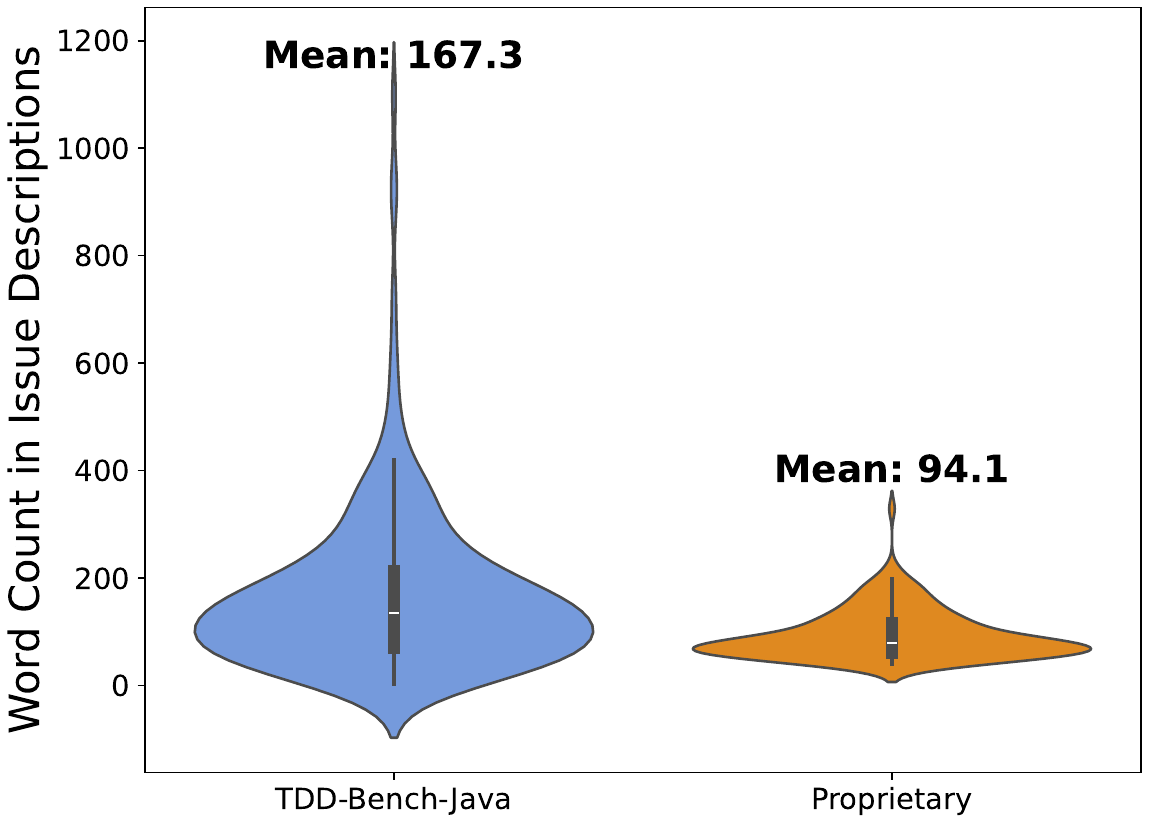}}
  \caption{\label{fig:word_count} Comparing the word count of issue descriptions in public TDD-Bench-Java vs.\ proprietary data.}
\end{figure}

We collected 150 issues from an actively-used and important IBM-internal
Java project and applied our proposed reproduction test generator.
In our public benchmark, all instances include test patches; however, in our proprietary dataset, only 28\% of instances 
have developer-written golden code patches. 
This suggests that developers may not always write reproduction tests. 
From our initial observations, issue descriptions in our proprietary data are much shorter than 
those in our public benchmark and often lack detailed information. 
\cref{fig:word_count} shows that the average word count for
issue descriptions~$d_\textrm{issue}$ in the public benchmark is 163.7, 
whereas for our proprietary dataset it is 94.1.
Additionally, the proprietary issue descriptions rarely 
contain code snippets or stack traces, making them more difficult to interpret compared to issues in TDD-Bench-Java.

\vspace{.2cm}
\noindent \textbf{Comparing code patches from public vs.\ proprietary data:} We compare open-source and proprietary code 
patches to estimate the difficulty of the issues. 
\cref{fig:1} shows the number of lines changed in code patches; the average number of lines changed is much higher for 
proprietary code patches (362.2 vs.\ 54.8). We also compare the number of diff-hunks in~\cref{fig:4}, total number of files (modified + new) in~\cref{fig:2}, and newly added files in~\cref{fig:3}. 
On average, 2.6 new files are added in our proprietary data, whereas there are hardly any new files added in public TDD-Bench-Java instances.
New files imply that new classes and functions are introduced, which are not available in the current repository~$c_\textrm{old}$.
Such newly-added classes or functions are difficult to anticipate for models,
making reproduction test generation more challenging. However, this raises the question of why code patches are significantly 
larger in proprietary projects compared to open-source projects. To answer this, we attempt to categorize the instances below.

\vspace{.2cm}
\noindent \textbf{Issue types: bug fix or feature request:} Developers write issues 
primarily for two reasons: (i) fixing a bug in the repository and (ii)~requesting a new feature. 
Adding a new feature typically involves more code changes compared to fixing bugs. 
LLMs have been shown to be effective at annotating software engineering artifacts~\cite{ahmed2025can, wang2025can}. 
We present the LLM with the issue description and corresponding code patch and 
ask it to categorize the issue as either a bug fix or a feature request. 
\cref{tab:open-pro} shows that 70.8\% of issue descriptions in TDD-Bench-Java are related to bug fixing, 
whereas in our proprietary data, only 16.0\% are bug fixes. In contrast, 29.2\% of issues in TDD-Bench-Java
are feature requests, 
which is much lower than in our proprietary project~(84\%). 
Since most issue descriptions in our proprietary data are feature requests, 
this helps explain the larger code patches.

\vspace{.2cm}
\noindent \textbf{Alignment between issue descriptions and code patches:} We are also interested 
in examining whether the code patches address the issue descriptions. 
Similar to the bug/feature categorization, 
we present the LLM with the issue description 
and corresponding code patch and ask whether the patch addresses the issue. 
\cref{tab:open-pro} shows that, for TDD-Bench-Java, the code patches 
align with the issue descriptions in 98.4\% of instances, 
whereas for our proprietary data, this alignment is observed in 81.3\% of cases.
This is in part due to the quality filters of public benchmarks and
further highlights the increased difficulty and challenges associated
with proprietary data.

\vspace{.2cm}
\findings{4}{Our proprietary code patches are significantly larger than those in our public TDD-Bench-Java, but the issue descriptions themselves are shorter and under-specified.}

\subsection{Performance of e-Otter on Proprietary Data}
\begin{figure}
  \centerline{\includegraphics[width=\columnwidth]{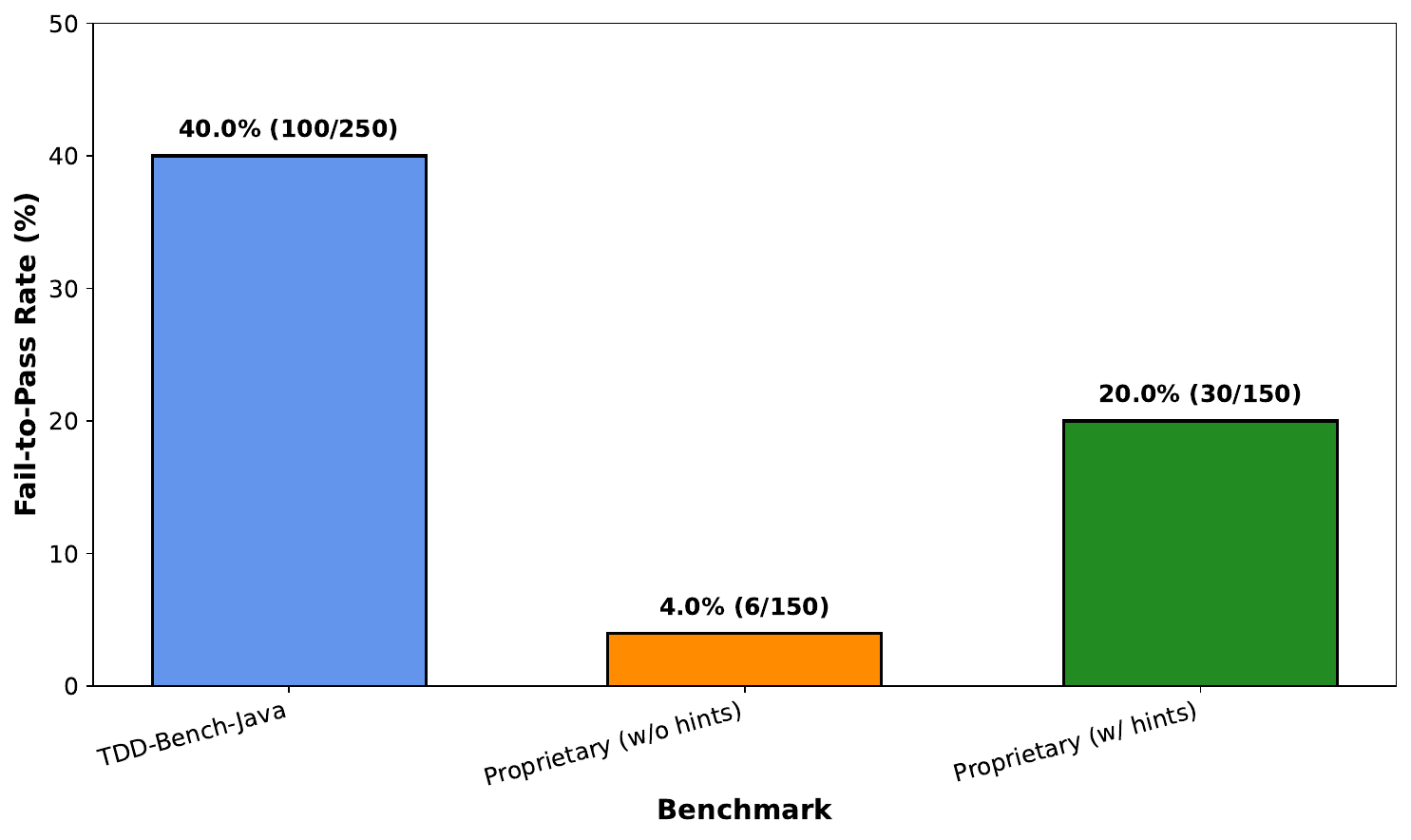}}
  \caption{\label{fig:compare} e-Otter performance on Open- and Closed-sourced Projects using Claude-Sonnet-4.5 model.}
\end{figure}

We ran e-Otter on our proprietary data and were able to generate fail-to-pass tests for only 
4\% (6 out of 150) of the instances~(\cref{fig:compare}), 
which is significantly lower than what we observed on TDD-Bench-Java. 
As discussed in~\cref{Characteristics}, our proprietary data often introduces new files and 
classes for each issue, making it difficult for the model to 
predict the correct class or function names, 
leading to syntactical errors. Even with execution feedback, 
we were unable to generate a sufficient number of fail-to-pass tests. Due to organizational policy, 
we used Claude-Sonnet-4.5 for these experiments.

One workaround for this problem is to add identifier
hints~\cite{deng_et_al_2025_nocode_bench} to the issue description,
including new class and function names. In real development settings, developers can provide such hints. 
In this paper, we present the code patch to the LLM and ask it to generate 
these hints. We then reran e-Otter and were able to fix 20\% of the instances (30 out of 150). 
Since the default issue descriptions are relatively short and 
performance is highly dependent on the hints, 
morphing primarily affects the hints rather than the issue description. 
Therefore, we did not apply inference scaling in this setup. 
However, we believe that improving the quality of hints can significantly enhance performance.

\vspace{.2cm}
\findings{5}{With the default issue description, e-Otter achieved only 4\% fail-to-pass rate on proprietary projects. 
However, adding hints with newly introduced class names and function signatures increased the performance to 20\%.
}

\subsection{Lession Learned and Future Direction}

When moving from open-source to proprietary projects, 
we encountered new challenges that can guide the design of better 
approaches in the future. First, the specifications or issue descriptions 
in proprietary projects are often not well written. 
Including hints can help in this scenario; 
however, several new concerns arise, such as how to generate better 
hints or improve the specifications. 
We can consider designing automated processes or incorporating human-in-the-loop approaches. 
Improving hints or specifications is an important problem, which we leave for future research.

\cref{tab:open-pro} shows how the performance of e-Otter changes between
our public TDD-Bench-Java and our proprietary dataset across different categories of issue descriptions. 
Interestingly, for bug fixes, performance differs little
between our public and private datasets (with hints):
for both, the fail-to-pass rate is around 45\%. 
However, we should interpret these results with caution, as the number of bug-fixing issues is relatively small in our proprietary dataset.
As expected, the fail-to-pass rate is higher for aligned samples.

\vspace{.2cm}
\findings{6}{Improving the issue description or specification using automated approaches or human-in-the-loop methods can be a promising direction for future work.
}

\begin{table}[t]
\centering
\caption{Performance of e-Otter on open-source and proprietary data across different categories}
\label{tab:open-pro}
%\vskip 0.05in
\resizebox{.85\columnwidth}{!}{%
\renewcommand{\arraystretch}{1.2}% Tighter
\begin{tabular}{llrrrr}
\toprule    
\multicolumn{1}{c}{\multirow{2}{*}{Type}} & \multicolumn{1}{c}{\multirow{2}{*}{Source}} & \multicolumn{2}{c}{Total}                             & \multicolumn{2}{c}{Fail-to-pass Test}                 \\
\multicolumn{1}{c}{}                      & \multicolumn{1}{c}{}                        & \multicolumn{1}{c}{Count} & \multicolumn{1}{c}{in \%} & \multicolumn{1}{c}{Count} & \multicolumn{1}{c}{in \%} \\ \midrule

\multirow{2}{*}{Bug}        & Open-source             & 177         & 70.8        & 78                & 44.1              \\
                            & Proprietary             & 24          & 16.0          & 11                & 45.8              \\ \cline{2-6}
\multirow{2}{*}{Feature}    & Open-source             & 73          & 29.2        & 22                & 30.1              \\
                            & Proprietary             & 126         & 84.0          & 19                & 15.1              \\ \midrule
\multirow{2}{*}{Aligned}    & Open-source             & 246         & 98.4        & 100               & 40.7              \\
                            & Proprietary             & 122         & 81.3        & 26                & 21.3              \\ \cline{2-6}
\multirow{2}{*}{Misaligned} & Open-source             & 4           & 1.6         & 0                 & 0.0                 \\
                            & Proprietary             & 28          & 18.7        & 4                 & 14.3          \\ \bottomrule         
\end{tabular}
}
\end{table}

\section{Discussion and Further Ablation}

\subsection{Impact of Iteration and Prior Change}
% In~\cref{tab:refiner}, we show the ablation for different prompts and how they contribute to \failtopass @ N. 
The refiner iteratively uses execution feedback and feeds the changes made in prior attempts to repair the test. 
\cref{tab:ablationeotter} shows that if we stop after the first execution feedback and do not try again, the performance drops by 15.0\% and 15.2\% for the Claude-Sonnet-4.7 and GPT-5.2 models.
Similarly, if we do not use the changes to prevent the model's repetitive errors, 
the performance drops by 12.0\% and~9.1\%. Therefore, both iteration and the summary of changes from prior attempts help improve performance.

\begin{table}[t]
\centering
\caption{Impact of Iteration \& Prior Change}
\label{tab:ablationeotter}
%\vskip 0.05in
\resizebox{.95\columnwidth}{!}{%
\renewcommand{\arraystretch}{1.2}% Tighter
\begin{tabular}{llrrr}
    \toprule
\multicolumn{1}{c}{Model}          & \multicolumn{1}{c}{Component} & \multicolumn{1}{c}{\failtopass} & \multicolumn{1}{c}{in \%} & \multicolumn{1}{c}{Change in \%} \\ \midrule
\multirow{3}{*}{Claude-Sonnet-4.5} & e-Otter                       & 100                     & 40.0                        & NA                               \\
                                   & e-Otter - iteration           & 85                      & 34.0                        & -15.0                              \\
                                   & e-Otter - change              & 88                      & 35.2                      & -12.0                              \\ \midrule
\multirow{3}{*}{GPT-5.2}           & e-Otter                       & 99                      & 39.6                      & NA                               \\
                                   & e-Otter - iteration           & 84                      & 33.6                      & -15.2                            \\
                                   & e-Otter - change              & 90                      & 36.0                        & -9.1     \\ \bottomrule                        
\end{tabular}
}
\end{table}

\subsection{Performance of Test Selector}
The test selector calls an LLM to choose the best candidate from the pool of test candidates. 
We divide the instances into three groups: (i) all six tests go from fail-to-pass, 
(ii) none of the tests go from fail-to-pass, and 
(iii) at least one test goes from fail-to-pass (but not all six). 
In the first two groups, it does not matter what the selector chooses: in the first group, 
the judge will always find a fail-to-pass test, and in the second group, the selector will never find one.
Only in the third group can the judge make a meaningful contribution. 
\cref{tab:testselection} shows the distribution of samples across these three groups and the selector's performance. 
On group (iii) where some but not all tests are \failtopass, we observe 64.1\% and 73.8\% accuracy for the selector for the Claude-Sonnet-4.5 and GPT-5.2 models, respectively.
% It means the Claude-Sonnet-4.5 and GPT-5.2 model-based selectors can find a fail-to-pass test with the reported accuracy.

\begin{table}[t]
\centering
\caption{Impact of Test Selector}
\label{tab:testselection}
%\vskip 0.05in
\resizebox{.95\columnwidth}{!}{%
\renewcommand{\arraystretch}{1.2}% Tighter
\begin{tabular}{lcrcr}
    \toprule
\multicolumn{1}{c}{\multirow{2}{*}{Model}} & \multicolumn{1}{c}{\multirow{2}{*}{Group}} & \multicolumn{1}{c}{\multirow{2}{*}{\# of Sample}} & \multicolumn{2}{c}{Selector's Success}             \\
\multicolumn{1}{c}{}                       & \multicolumn{1}{c}{}                       & \multicolumn{1}{c}{}                              & Count                  & \multicolumn{1}{c}{in \%} \\ \midrule
\multirow{3}{*}{Claude-Sonnet-4.5}         & All (6) \failtopass                                    & 59                                                & \multicolumn{2}{c}{NA}                             \\
                                           & None (0) \failtopass                                   & 113                                               & \multicolumn{2}{c}{NA}                             \\
                                           & Some (1-5) \failtopass                                    & 78                                                & \multicolumn{1}{r}{50} & 64.1                      \\ \midrule
\multirow{3}{*}{GPT-5.2}                   & All (6) \failtopass                                    & 57                                                & \multicolumn{2}{c}{NA}                             \\
                                           & None (0) \failtopass                                   & 113                                               & \multicolumn{2}{c}{NA}                             \\
                                           & Some (1-5) \failtopass                                    & 80                                                & \multicolumn{1}{r}{59} & 73.8        \\ \bottomrule              
\end{tabular}
}
\end{table}

\subsection{Cost Estimation}

In e-Otter++, we generate six test candidates and use a selector at the end to choose the best ones. 
It costs us approximately \$2.5 and \$1.5 for each instance with the Claude-Sonnet-4.5 and GPT-5.2 models, respectively. 
Thus, for each candidate, it costs \$0.42 and \$0.30. We can reduce or adjust the number of candidates to control overall spending.

\section{Threats to Validity}

In prior work, coverage has been considered as an additional evaluation metric. 
We observe that fail-to-pass tests tend to have higher coverage on average than non-fail-to-pass tests. 
However, many non-fail-to-pass tests exhibit similar coverage levels.
It makes coverage an unreliable indicator of the fail-to-pass property of the test. 
As a result, we do not consider coverage as a primary metric for reproduction test evaluation. 
Additionally, while Python allows lightweight coverage measurement without modifying repositories, 
Java tools such as JaCoCo require bytecode instrumentation and additional configuration. 
Since our primary objective is to evaluate fail-to-pass behavior, 
we avoid incorporating JaCoCo to keep the evaluation harness lightweight, reduce complexity, and prevent potential side effects. 
This also ensures consistent evaluation across repositories with different build systems and configurations.

One of the key concerns with this line of generative work is the contamination or memorization problem of the model~\cite{badertdinov_et_al_2025,liang_garg_moghaddam_2025}. 
However, in this paper, we evaluated e-Otter++ on proprietary data, which gives us a contamination-free evaluation. 
From our experiments, it is evident that models struggle with unseen repositories, 
but we also observe challenges specific to the proprietary issues. 
Therefore, it is not conclusive how much model performance is influenced by model contamination versus issue difficulty.

We find that resolving issues in proprietary projects is more difficult than in open-source projects. 
However, our findings are limited to our proprietary dataset of 150 issues 
and may not generalize to other proprietary projects within our organization or elsewhere.

\section{Related Work}\label{sec:related}

\paragraph{Benchmarks for Java reproduction test generation.}
Our work on TDD-Bench-Java is inspired by recent 
repository-level software engineering benchmarks such as
SWE-bench~\cite{jimenez_et_al_2024} and SWE-bench
Verified~\cite{chowdhury_et_al_2024}.
Those initial benchmarks focused on the patch generation task for
Python, and were soon followed by benchmarks for the repository-level
reproduction test generation task, specifically,
SWT-bench~\cite{mundler_et_al_2024},
TDD-Bench-Verified~\cite{ahmed_et_al_2024}, and later
USEBench~\cite{applis_et_al_2026}.
However, none of the above benchmarks cover Java.
Repository-level software engineering benchmarks that cover Java include 
Defects4J~\cite{just_jalali_ernst_2014},
SWE-PolyBench~\cite{rashid_et_al_2025}, and
Multi-SWE-bench~\cite{zan_et_al_2025}.
Unfortunately, the task in these benchmarks is not reproduction test
generation.
A recent benchmark, OmniCode~\cite{sonwane_et_al_2026}, includes
repository-level Java instances and a ``test generation'' task.
However, that task does not refer to reproduction tests: it does not
evaluate the test on the old pre-patch code~$c_\textrm{old}$, which
would be required to reproduce the issue.
Thus, TDD-Bench-Java is the first repository-level Java benchmark for
reproduction test generation.

\paragraph{Solutions for Java reproduction test generation.}
Our work on Otter for Java is inspired by recent agents and workflows
for repository-level software engineering tasks such as
SWE-Agent~\cite{yang_et_al_2024},
Agentless~\cite{xia_et_al_2025}, and
OpenHands~\cite{wang_et_al_2025_openhands}.
Those initial solutions focused on the patch generation task for
Python, and were soon followed by agents and workflows for
repository-level reproduction test generation, such as
Otter~\cite{ahmed_et_al_2025},
Aegis~\cite{wang_et_al_2024},
e-Otter~\cite{ahmed_et_al_2026}, and
AssertFlip~\cite{khatib_mathews_nagappan_2026}.
However, none of the above solutions cover Java.
Repository-level agents and workflows that cover Java include 
InfCode~\cite{li_et_al_2025_infcode},
Prome\-theus~\cite{chen_et_al_2025}, and
iSWE~\cite{ganhotra_et_al_2026}.
Unfortunately, the task they solve is not reproduction test generation.
There are very few solutions for Java reproduction test generation:
Libro~\cite{kang_yoon_yoo_2023} is a workflow of two LLM calls, and
BRT Agent~\cite{cheng_et_al_2025} is an agentic solution.
However, neither of these is evaluated in a setting where they
perform their own localization.
Thus, Otter for Java is the first complete solution for
repository-level Java reproduction test generation.

\paragraph{Industry experience for Java reproduction test generation.}
A serious threat to validity pervading the research literature on
repository-level software engineering is that open-source repositories
may not be representative of industry code.
One concern is that large language models tend to be trained on huge
datasets that include all available open-source software, which means
that their training data is (at least to some extent)
contaminated~\cite{chen_ma_jiang_2025}.
This has given rise to benchmarks such as
SWE-rebench~\cite{badertdinov_et_al_2025}, which resists contamination
by releasing new instances predated by LLM training data cut-offs,
and SWE-Bench Pro~\cite{deng_et_al_2025}, which resists contamination
by using data whose license prohibits use for LLM training.
Unfortunately, neither of these benchmarks covers Java, and neither of
them evaluates the reproduction test generation task.
Since our new TDD-Bench-Java benchmark leverages public data, this
paper additionally also reports insights based on a proprietary
industry dataset.
In doing so, it takes inspiration from recent papers on
BRT Agent with insights on Google-internal code~\cite{cheng_et_al_2025};
HULA with insights on Atlassian-internal code~\cite{takerngsaksiri_et_al_2025};
and AutoCodeRover with insights on SonarSource-internal
code~\cite{mirchev_et_al_2026}.
Of these, only BRT Agent focuses on reproduction test generation, and
the number of Java instances involved is much smaller than in our
experiments.

\section{Conclusion}\label{sec:conclusion}

Issue reproduction tests are tests that demonstrate the presence of
an open issue before and its resolution after it gets fixed.
These tests serve several core software engineering needs including
requirements clarification, issue resolution, and quality assurance
via continuous integration.
Unfortunately, while there have been some benchmarks and solutions for
reproduction test generation for Python, they have been missing for
Java.
This paper takes steps to remedy that situation by introducing a new
public benchmark~(TDD-Bench-Java), and by describing and evaluating an
LLM-based workflow for it~(Otter for Java).
Since results on carefully curated public benchmarks do not always
generalize to an industrial setting, this paper complements the
experiments on public data with experiments on proprietary data.
Not surprisingly, that setting is difficult and requires more
attention and innovation.

\vspace{.3cm}
\noindent \textbf{Data Availability Statement:} The code for the reproduction test generators is proprietary. 
We have open-sourced TDD-Bench-Java at \textcolor{blue}{\href{https://github.com/IBM/TDD-Bench-Verified/tree/main/TDD-Bench-Java}{TDD-Bench-Java Repository}}.
and shared our prompts at \textcolor{blue}{\href{https://drive.google.com/drive/folders/1i6Tp2AvlN7x7lWcGw_npeZIZvGGWleWl?usp=sharing}{Link}}.   

%\newpage

\balance
\bibliographystyle{ACM-Reference-Format}
\bibliography{bibfile}

\end{document}